\documentclass[
    twocolumn,
	prd,
	amssymb,
	preprintnumbers,
	secnumarabic,
	nofootinbib]{revtex4-1}

\pdfoutput=1

\usepackage{inputenc}
\usepackage{graphicx}
\usepackage{enumitem}
\usepackage{latexsym}
\usepackage{amsfonts}
\usepackage{amssymb}
\usepackage{color}
\usepackage{amsmath}
\usepackage{slashed}
\usepackage{dcolumn}
\usepackage{verbatim}
\usepackage{float}
\usepackage{multirow}
\usepackage{xspace}
\usepackage[normalem]{ulem}
\usepackage[
pdfauthor=Nirmal Raj]{hyperref}

\newcommand{\gsim}{\gtrsim}

\newcommand{\ra}{\rightarrow}

\def\Oc{\mathcal{O}}

\newcommand{\acro}[1]{\textsc{\MakeLowercase{#1}}} 
\newcommand{\osn}{\oldstylenums}
 
\newcommand{\beq}{\begin{equation}}
\newcommand{\eeq}{\end{equation}}
\newcommand{\bea}{\begin{eqnarray}}
\newcommand{\eea}{\end{eqnarray}}
\newcommand{\nn}{\nonumber}

\def\mdm{m_\chi}
\def\sigmanDM{\sigma_{{\rm n}\chi}}

\definecolor{orange}{rgb}{1,0.5,0}
\definecolor{purple}{rgb}{1,0,1}
\definecolor{brown}{rgb}{.7,.2,.2}
\definecolor{violet}{rgb}{.6,.3,.8}
\definecolor{nicegreen}{rgb}{.3,.7,.3}

\allowdisplaybreaks

\begin{document}

\title{Dark matter astrometry at underground detectors with multiscatter events }

\author{Joseph Bramante}
\affiliation{The Arthur B. McDonald Canadian Astroparticle Physics Research Institute, Department of Physics, Engineering Physics,
and Astronomy, Queen's University, Kingston, Ontario, K7L 2S8, Canada}
\affiliation{Perimeter Institute for Theoretical Physics, Waterloo, ON N2J 2W9, Canada}

\author{Jason Kumar}
\affiliation{Department of Physics and Astronomy, University of Hawaii, Honolulu, Hawaii 96822, USA}

\author{Nirmal Raj}
\affiliation{TRIUMF, 4004 Wesbrook Mall, Vancouver, BC V6T 2A3, Canada}

\begin{abstract}

We show that current and imminent underground detectors are capable of precision astrometry of dark matter.
First we show that galactic dark matter velocity distributions can be obtained from reconstructed tracks of dark matter scattering on multiple nuclei during transit;
using the liquid scintillator neutrino detector \acro{sno}+ as an example, we find that the dark matter velocity vector can be reconstructed event-by-event with such a small uncertainty, that the precision of dark matter astrometry will be limited mainly by statistics.
We then determine the number of dark matter events required to determine the dispersion speed, escape speed, and velocity anisotropies of the local dark matter halo, and also find that with as few as $\Oc(10)$ events, dark matter signals may be discriminated from potential backgrounds arising as power-law distributions.
Finally, we discuss the prospects of dark matter astrometry at other liquid scintillator detectors, dark matter experiments, and the recently proposed \acro{mathusla} detector.
\end{abstract}

\maketitle

\section{Introduction}

While the search for particle dark matter colliding at most once per transit through underground detectors has proceeded apace, the study of dark matter scattering multiple times has recently received reinvigorated research.
It has been shown that dark matter candidates expected to interact multiple times in detectors can be discovered using entirely new analyses at both traditional single-scatter dark matter experiments and neutrino detectors~\cite{1803.08044,1812.09325}.
Multiply interacting dark matter candidates include electroweak-symmetric solitons~\cite{1906.10739},
baryon-charged particles~\cite{1809.07768},
composites with \acro{QCD}-charged constituents~\cite{Kang:2006yd,0712.2681,1801.01135}
dark nucleons~\cite{1812.07573},
light mediator models~\cite{1812.09325}, and
primordial charged black holes~\cite{Lehmann:2019zgt}.
In this work, we focus on the prospects for such detectors to use dark matter velocity data to discriminate candidate dark matter events from backgrounds, and follow up any
discovery with precision astrometry of dark matter.

The principle underlying our study is that one can effectively reconstruct the track of a source of scintillation light within a liquid
scintillator detector from the timing and location of the photomultiplier tube (\acro{PMT}) illuminations.
This idea was implemented in an analysis of KamLAND data, and was used
to reconstruct the track of charged particles produced by neutrino charged current interactions~\cite{
0902.4009,
0908.1768,
0909.4974,
1103.3270,
Sakai:2016plr}.
Our strategy will be to similarly reconstruct the track of a
dark matter particle traversing the inner detector, utilizing the fact that the heavy dark matter particle essentially transits the detector in a straight line at
constant speed, with scintillation light originating from the points along this track where much lighter nuclei
recoil.
Although this strategy is similar to previous efforts involving leptons produced by charged current events, there are two key
distinctions: one, the dark matter tracks we construct will generate fewer \acro{PMT} hits, and two, these hits will be more widely spaced in time since dark matter moves more slowly.
With these distinctions, we can make reasonable estimates for the precision with which we can reconstruct a dark matter particle track.

Besides providing a simple method to validate candidate dark matter events against possible backgrounds, this effort may ultimately yield a measurement of the dark matter velocity distribution in the vicinity of the Solar System.
Such a measurement would not only be relevant to understanding dark matter astrophysics, but would allow one to distinguish a putative dark matter
signal from possible backgrounds.
There is, of course, a large experimental effort in directional dark matter direct detection with the similar goal of distinguishing backgrounds
and performing dark matter astrometry~\cite{directional}.
This effort
focuses on reconstructing the direction and energy of a nucleus recoiling from a dark matter scattering event; the direction and energy of the dark matter
particle itself is unmeasured, and can only be inferred statistically from a large number of events, under some assumptions regarding the nature of the
dark matter-nucleus interaction.
In contrast, our strategy allows one to measure the direction and speed of a dark matter particle directly, on an event-by-event
basis.
This provides a much more precise and model-independent tool for dark matter astrometry.

We also determine how many dark matter events are required, assuming reasonable estimates for reconstruction uncertainties, to distinguish between different
models of the dark matter speed distribution, to distinguish velocity distribution models which are isotropic in the galactic frame from anisotropic ones, and to
distinguish a possible dark matter signal from a possible instrumental, radiogenic, or cosmogenic background.  We will also find that we can use dark matter to perform ``non-dark" astronomy,
by providing a precise measurement of the galactic escape speed.
This is a unique measurement; essentially, one would be directly measuring the speeds of
gravitationally-interacting particles which are the most weakly bound to the Milky Way halo.
As we will show later, our method of  dark astrometry may determine the galactic escape speed to within a precision comparable to, or even better than, current astrophysical surveys.

The plan of this paper is as follows.
In the following section, we first sketch the cross sections and dark matter masses that can be probed at the liquid scintillator-based \acro{sno}+ experiment. 
This provides an estimate of the number of multiscattering dark matter events that may be collected, and the maximum interaction length of dark matter.
Next we detail the event-by-event reconstruction of the dark matter velocity vector, and estimate the attendant uncertainties.
We then use this information to extract key kinematic properties of the speed distribution, to reject environmental backgrounds that may arise as power-law distributions, and to determine anisotropies in the dark matter angular distribution, commenting on the effect of statistics and detector resolutions.
In Sec~\ref{sec:conc}, we discuss the prospects of dark matter astrometry at other other neutrino detectors, dark matter experiments and the \acro{lhc}-based \acro{mathusla} detector, and then conclude.

\begin{figure}
\includegraphics[width=.45\textwidth]{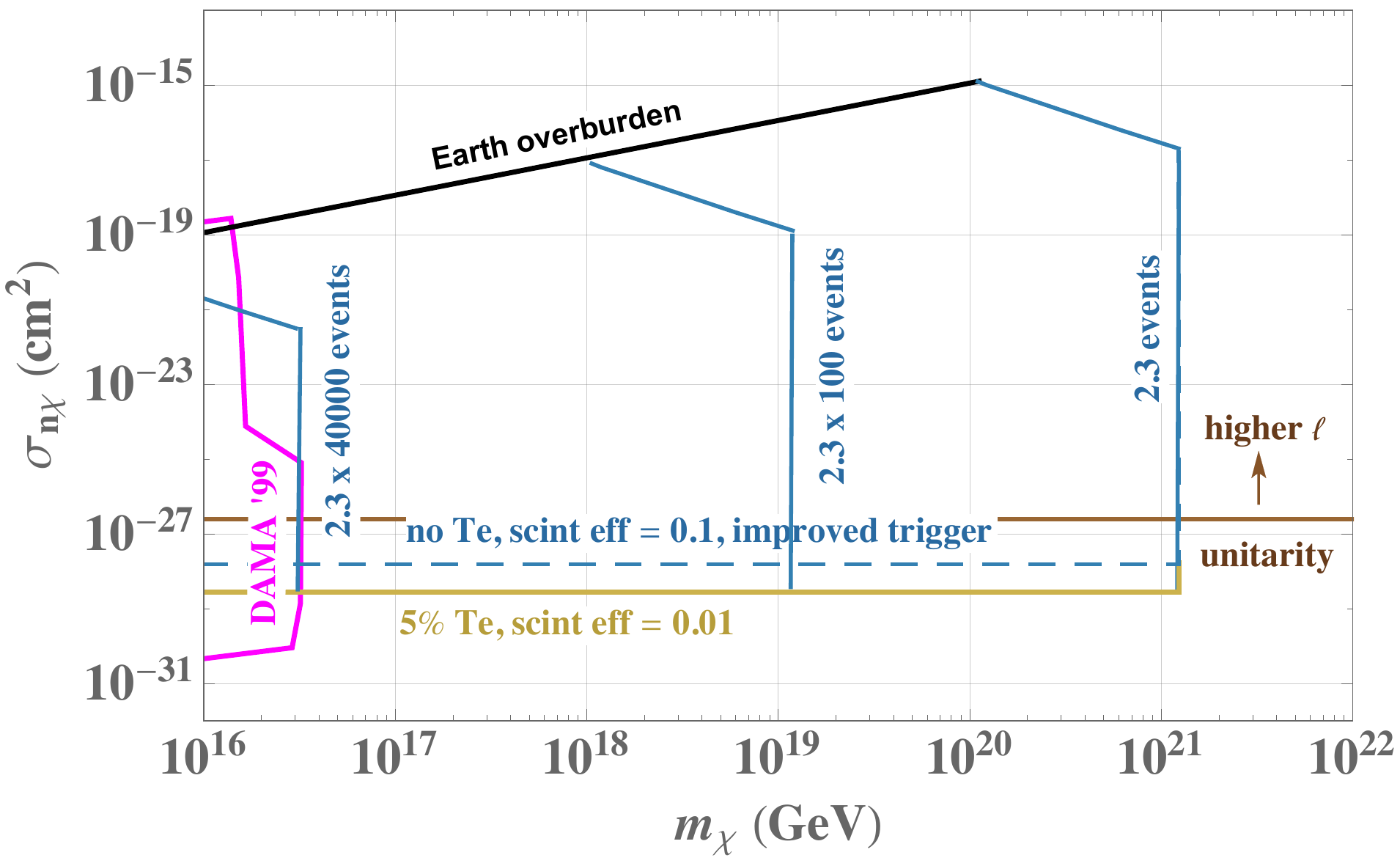}
\caption{
The 10-year reach of \acro{sno}+ in a search for spin-independent multiscattering dark matter.
Also shown are the improvement in cross section sensitivity gained with 5\%-by-weight loading of $^{130}$Te in the liquid scintillator;
the cross section above which $s$-wave perturbative unitarity is violated;
the cross sections above which the Earth's overburden would degrade dark matter's energy below detector thresholds; and the erstwhile constraints placed by \acro{dama}.
S\acro{NO}+ could detect $\leq 10^5$ dark matter events, acquiring excellent statistics to perform dark matter astrometry.
}
\label{fig:reach}
\end{figure}

\section{Sensitivity}

Even though our general ideas apply to all types of detectors, for demonstration we analyse a concrete example: the liquid scintillator neutrino experiment \acro{sno}+ situated in SNOLab.
In Sec.~\ref{sec:conc} we discuss some specific ideas for other experimental configurations.

\subsection{Reach in cross sections and dark matter masses}

In Fig.~\ref{fig:reach} we show the 10-year-runtime reach of \acro{sno}+ in the space of per-nucleon cross section $\sigmanDM$ vs dark matter mass $\mdm$ for the case of spin-independent scattering, along with the existing bound from a multiscatter search at \acro{dama}~\cite{Bernabei:1999ui}.
The \acro{sno}+ reach in $\sigmanDM$ is determined by the minimum number of photoelectrons detectable within a timing window as dark matter transits the detector,
and the reach in $\mdm$ is determined by the integrated dark matter flux admitted by the detector over the run-time; see Ref.~\cite{1812.09325} for further details.

S\acro{no}+ would in the near future search for neutrinoless double beta decay of tellurium ($^{130}$Te) loaded into the liquid scintillator \acro{LAB} (linear alkyl benzene).
This may advantage a search for spin-independent multiscattering dark matter since  the scattering would be coherently enhanced by the large nucleon number of $^{130}$Te.
In Fig.~\ref{fig:reach} we show the reach in $\sigmanDM$ for 5\% Te-loading by weight and compare it to the reach with pure \acro{lab}.
Extrapolating from results in Ref.~\cite{Hong:2002ec}, we have factored in a scintillation efficiency (the light yield per recoil) of 0.01 and 0.1 for Te (recoiling at $\sim$ 100 keV) and C ($\sim$ 10 keV) targets respectively.

We have extrapolated the dark matter-tellurium scattering cross section from the dark matter-carbon cross section under the assumption that dark matter is a point-like particle, with equal coupling
to protons and neutrons, and whose amplitude for coherent scattering on nuclei is $s$-wave and perturbative.
These assumptions are invalid if the scattering cross section is sufficiently large~\cite{1812.09325,1907.10618}, in which case perturbative $s$-wave unitarity would be violated, unless the dark matter is a composite state or scatters via a long-range mediator.
We have indicated this $s$-wave unitarity limit with a horizontal brown line, obtained by setting the per-nuclear scattering cross section to $16\pi/s = 4\pi/(\mu_{\rm C \chi} v_\chi)^2$ for the partial wave $\ell=0$, where $\mu_{\rm C \chi}$ is the carbon-dark matter reduced mass and $v_\chi = 10^{-3}~c$.
For higher partial waves this limit is scaled up by a factor of 2$\ell$+1.
In the region above the line, the scattering cross section normalized to nucleons is too large to consistently treat the dark matter as point-like, and the dark matter-nuclear interaction as a contact interaction.
We note that the improvement in cross section reach does in fact occur in a regime where point-like, perturbative $s$-wave scattering {\em is} allowed; in this region, where the above assumptions are satisfied, our improved reach estimate is valid.

From the vertical lines in the plot we see that in the parameter space unconstrained by \acro{dama}, up to $N_{\rm DM}$~=~9.2~$\times 10^4$ dark matter events may be discovered by \acro{sno}+.
Thus one could hope for enough statistics to perform dark matter astrometry in case a dark matter search yields positive signals.
Accordingly, we consider scenarios where the number of dark matter tracks seen at \acro{sno}+ ranges from ${\cal O}(10^0-10^5)$.
Note that, although the scattering cross section determines whether or not a transiting dark matter particle will
deposit sufficient energy in the detector to exceed thresholds and leave a track, the mass of the dark matter particle (which
determines the dark matter number density) determines the event rate.  Thus, our subsequent analysis of dark matter
astrometry does not depend on whether
or not dark matter is a point particle or composite, or on whether or not dark matter-nucleus scattering can be extrapolated from
dark matter-nucleon scattering.
We need only assume that the dark matter-nucleus (either carbon or tellurium) scattering cross section
is sufficiently large that every transiting particle deposits an amount of energy which exceeds threshold, but not so much
energy that its speed is appreciably degraded.
The particle mass then
determines the number of events which will be seen, which in turn determines the precision with which we can measure dark matter
astronomical observables.

\subsection{Reconstructing dark matter speed and direction}
\label{sec:reconstruct}

 \begin{table}
 \begin{center}
\begin{tabular}{| c | c |}
  \hline			
  Variable uncertainty & Baseline resolution \\
  \hline
  angle: $\bf{\delta \psi}$ & $\mathbf{3.7 \times 10^{-2}}$ \\
  longitudinal path length: $\delta L/L$ & 6.7 $\times 10^{-4}$ \\
  timing: $\delta T/T$ & 10$^{-4}$  \\
  speed: $\bf{\delta v/v}$ &  $\mathbf{6.7 \times 10^{-4}}$ \\
  \hline
\end{tabular}
\end{center}
\caption{The resolutions of variables that may be reconstructed at \acro{sno}+ used as a baseline in this work.
See Sec.~\ref{sec:reconstruct} for how these are estimated.
In a real experiment, these uncertainties may vary event by event, but in this work we assume uniform resolution across event samples.
The small values of angular and speed uncertainties imply that smearing due to detector resolution will not substantially limit multiscattering event reconstruction.}
\label{tab:res}
\end{table}

We now show how the speed and direction of the velocity vector of dark matter may be reconstructed at \acro{sno}+, and estimate the associated uncertainties.
In practice these variables would be reconstructed in the detector frame, but for some of our analysis we will assume that these have been boosted back to the galactic frame.

We estimate the uncertainty on speed and direction by determining how accurately the \acro{PMT}s at \acro{sno+} can reconstruct the position and time at which dark matter particles enter and exit the inner detector.
For the moment assuming that \acro{PMT}s nearest to a nuclear recoil will register the most scintillation light, it follows that \acro{PMT}s nearest to the point where a particle enters the detector will light up first, and record a high number of photoelectrons.
Likewise, the \acro{PMT}s near the point of exit will light up last, also recording numerous photoelectrons.
All other \acro{PMT}s, relatively distant from the particle trajectory, would record fewer photoelectrons.
Thus the entry and exit points constitute two ``hot spots" that
allow us to reconstruct the particle's direction and path length through the inner detector; in addition, the time interval between the appearance of these hot spots gives the particle's speed.
The uncertainties associated with this reconstruction may be estimated as follows.

The angular uncertainty is given by
\bea
\delta \psi \simeq \frac{\Delta d}{L},
\eea
where $L$ is the reconstructed path length and $\Delta d$, the transverse uncertainty in $L$, is the maximum of the spacing between \acro{PMT}s, $\Delta d_{\rm PMT}$,
and the dark matter interaction length, $\lambda$.
As for the uncertainty $\delta v$ in speed $v$ for a detector transit time $T$, we have $v=L/T$, so that
\bea
\frac{\delta v}{v} &=& \sqrt{\left(\frac{\delta L}{L} \right)^2 + \left(\frac{\delta T}{T} \right)^2}~,
\eea
where $\delta T$ is the timing uncertainty and $\delta L$ is the longitudinal uncertainty in $L$.
Here $\delta T$ may be estimated as the maximum of the \acro{PMT} timing
resolution and the time it takes for a scintillation photon to travel between neighboring \acro{PMT}s, $\Delta d/(\kappa c)$, with $\kappa$ the refractive index of \acro{lab} scintillator taken to be 1.5~\cite{LABrefracindex}. 
The quantity $\Delta d/(\kappa c)$ is also the uncertainty in determining when the dark matter enters or exits based on when the \acro{PMT}s were lit.
Note that $\delta L$ is given by $(\Delta d)^2/2L$, so that $\delta L/L = (\delta \psi)^2/2$.

At \acro{SNO}+, 9300 \acro{PMT}s light guides surround a 6 meter-diameter inner detector. As discussed in \cite{1812.09325}, for multiscattering dark matter to be discovered at this experiment, we conservatively require a minimum of $\sim 100$ photons produced and detected by \acro{PMT}s during the $\sim 10~{\rm \mu s}$ transit of a multiscattering particle. 
This should be compared  to the expected dark count rate across the entire detector, which is $\sim 10$ \acro{PMT} dark counts in the same time period. With only ten dark counts recorded across 9300 \acro{PMT}s, and with at least $\sim 100$ photons recorded along the dark mater multiscattering track, we expect one signal photon to be produced in a dark matter recoil every $\sim 5~{\rm cm}$ or less, with little interference from dark counts. Therefore, we expect the limiting length scale for determining where the dark matter enters and exits the detector, is the separation between light guides, which is $\Delta d_{\rm PMT} = 0.11$~m for the 9300 \acro{PMT}s light guides surround the 6 meter-diameter inner detector at \acro{SNO}+.
In more detail, if we take the typical path length to be the inner detector radius, $L$ = 3~m, then from Fig.~\ref{fig:reach} we estimate that dark matter scattering on carbon can be detected if $\lambda$ = (carbon number density)$^{-1} \ \times$ (threshold cross section)$^{-1}$ $\leq 0.07$~m.
Thus $\Delta d$ = 0.11~m, from which
we also have $\delta \psi = 0.037$.
From the value of $\Delta d$, we see that a scintillation photon takes 0.6~ns to move between \acro{PMT}s.
The \acro{sno}+ timing resolution is expected to be around 1~ns\footnote{We thank Alex Wright for guidance on this point. This estimate also matches event time binning at \acro{BOREXINO}~\cite{BOREXINO}.}, hence $\delta T$ = 1 ns.
The typical dark matter transit time $T = L/(300 \ {\rm km \ s}^{-1}) = 10^{-5}$~s.
We also find that $\delta L = 2 \times 10^{-3}$ m.
Putting these together, we see that $\delta v/v = 6.7 \times 10^{-4}$.
Thus the uncertainty in reconstructing dark matter speed is $<$ 1 km/s, well below the characteristic dark matter speed in our local halo.
We will use the above uncertainties, summarized in Table~\ref{tab:res}, as a baseline in our following calculations.

\begin{figure*}
\includegraphics[width=.45\textwidth]{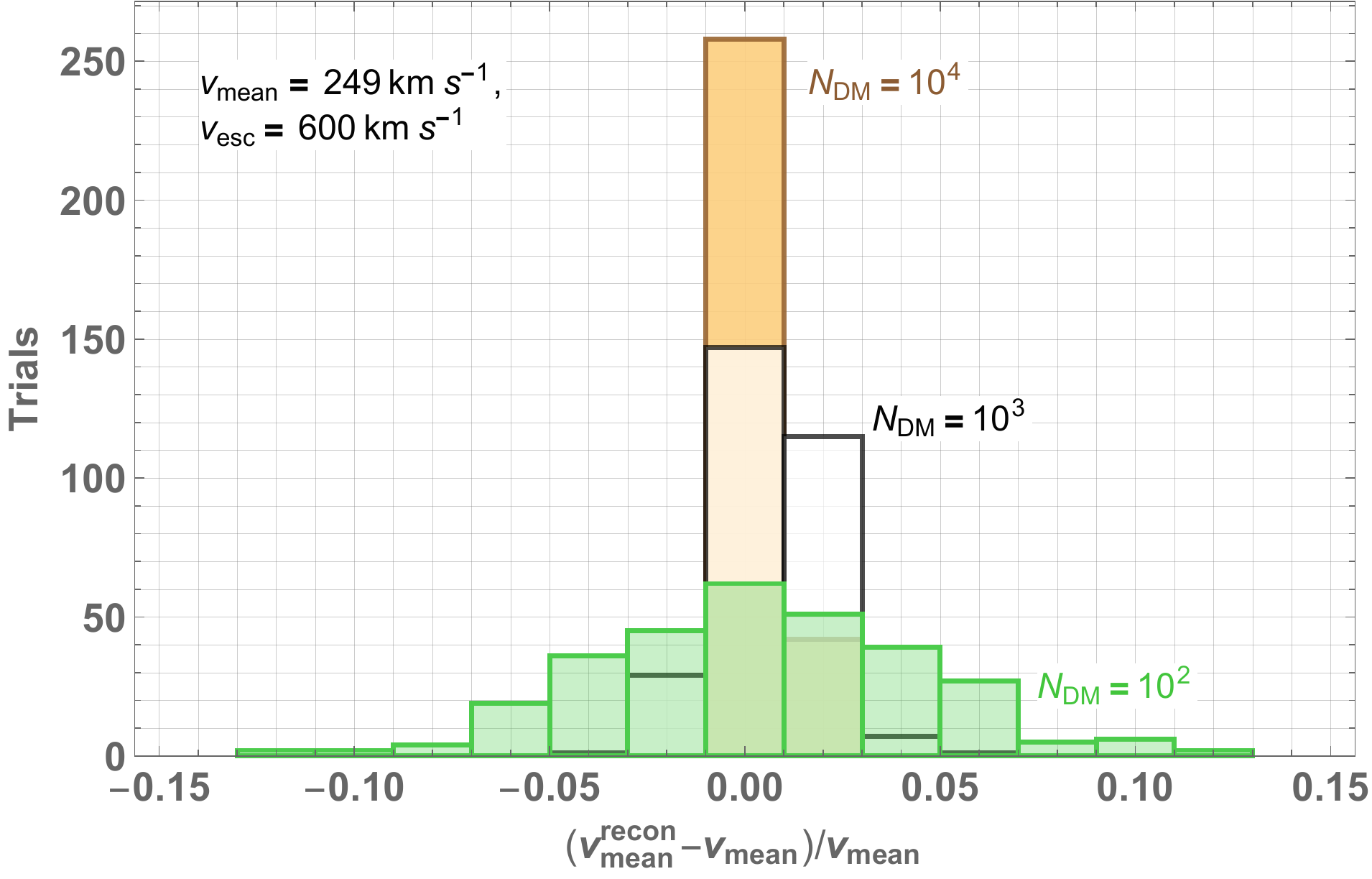}
\includegraphics[width=.47\textwidth]{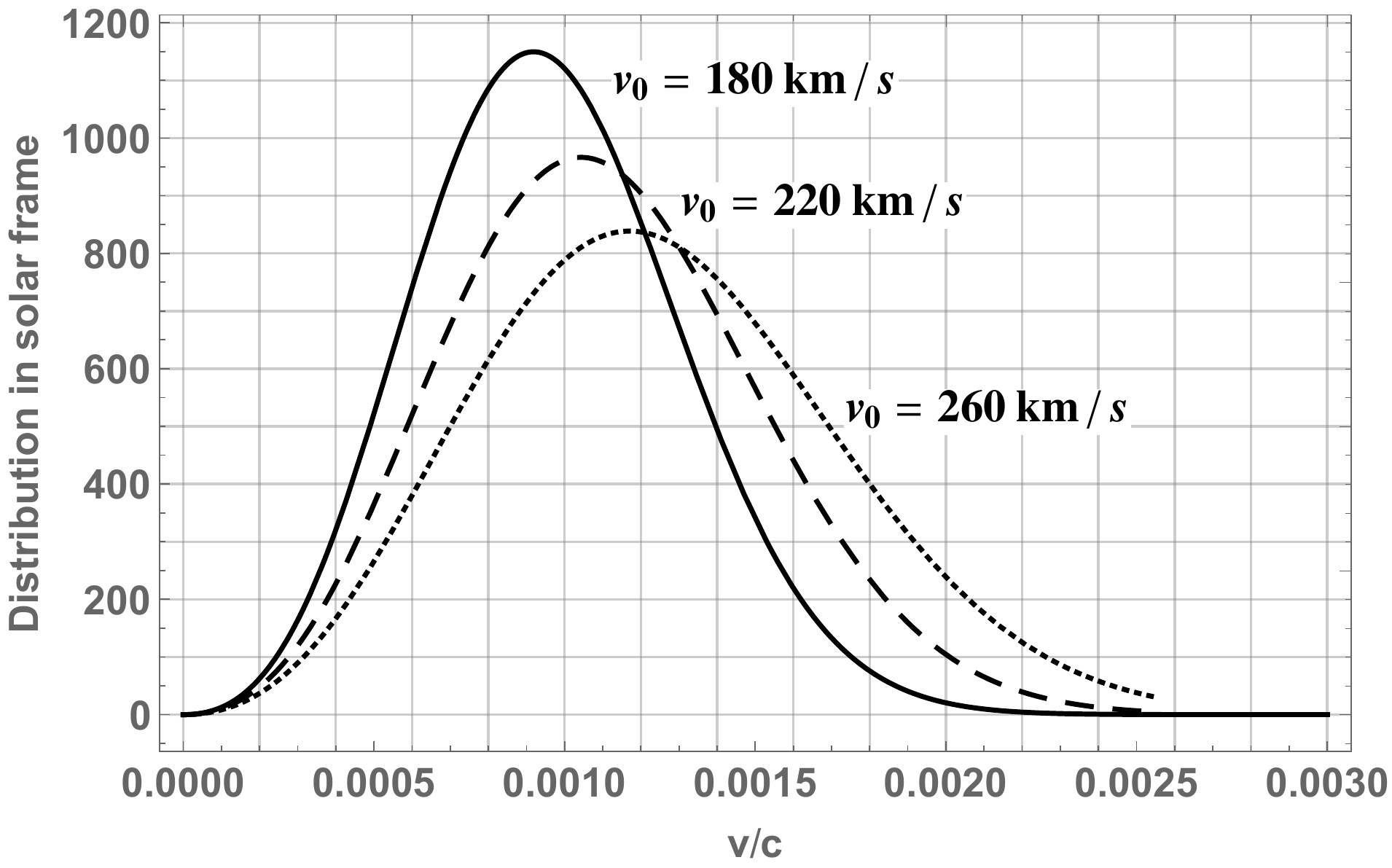} \\
\includegraphics[width=.94\textwidth]{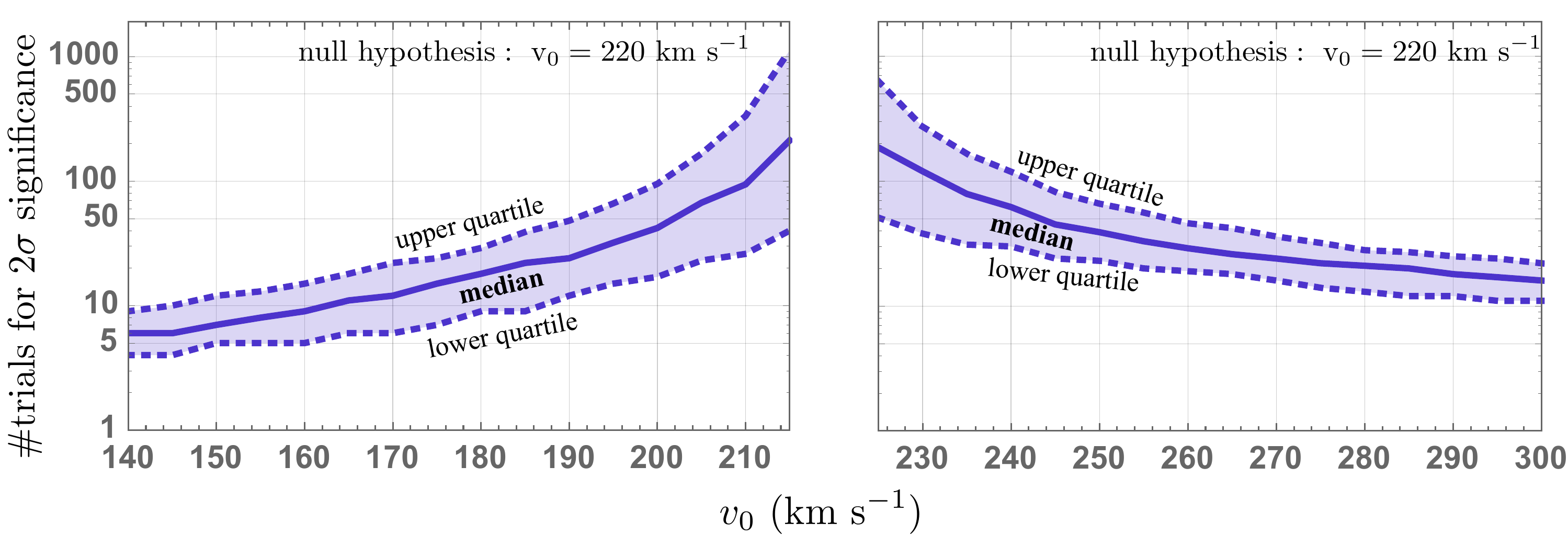}
\caption{
{\bf Top left}: Distribution over 300 trials of the normalized difference between the reconstructed-at-\acro{sno}+ and actual mean speeds of dark matter, with 10$^2$, 10$^3$, and 10$^4$ events.
As expected, greater statistics would help more precise reconstruction.
See Sec.~\ref{subsec:speedspec} for further details.
{\bf Top right}: Normalized dark matter halo distributions in the Earth's rest frame $f_\oplus(v)$ taken from Eq.~\eqref{eq:vear} with circular speeds $v_0 =180,220,260$ km~s$^{-1}$.
{\bf Bottom}: Results from 10$^3$ Kolmogorov-Smirnov trials determining the number of events required to distinguish a distribution with circular speed $v_0$ from that with $v_0 =220$ km~s$^{-1}$.
The dashed lines denote the upper and lower quartile of trials, the solid line denotes the median trial.
}
\label{fig:vdisp}
\end{figure*}

A more complete analysis would of course involve event reconstruction from all the data from all the \acro{pmt}s;
moreover, these uncertainties could vary from event to event, and could depend on any cuts imposed on the minimum path length, but we leave these tasks to the experimental collaborations, who are best equipped to carry them out.
For the purpose of a proof-of-principle estimate, we assume that the above-estimated uncertainties are uniform across events samples.
Our estimates above already show that \acro{sno}+ could perform dark matter astrometry with very high precision.
In particular, the small uncertainties in velocity and angle reconstruction suggested that smearing  of velocity distributions due to detector resolution should not be important, and this is indeed what we find in practice.
More specifically, we find that our baseline $\delta v/v$ is so small that only when this uncertainty is $\Oc$(100) larger does it induce any appreciable smearing; our baseline $\delta \psi$ is $\Oc$(1) short of inducing the same.
Thus we will only vary $\delta \psi$ when generating events.

In the following, we show that reconstructing the dark matter direction and speed as described above would enable us to reconstruct the speed and angular distribution of galactic dark matter, directly extract several kinematic properties of dark matter, and discriminate between signal and background distributions.

\subsection{Speed distribution extraction and background rejection}
\label{subsec:speedspec}

In this section we show that with enough multiscatter event statistics, the dark matter speed distribution may be reconstructed, and in particular important properties such as the speed dispersion and local halo escape speed may be estimated.
An empirical distribution of observed speeds can also be used to reject backgrounds that may arise as a power-law distribution.

We first pick our benchmark speed distribution as the Maxwell-Boltzmann distribution in the galactic frame:
\bea
 f(v) = \frac{1}{\mathcal{N}}
v^2
\exp\left(-\frac{v^2}{v^2_0}\right)
\Theta(v_{\rm esc} - v)~,
\label{eq:fMB}
\eea
where $\Theta$ is the Heaviside theta function and
the normalization factor is given by
\bea
\nn \mathcal{N} =
{\frac{\sqrt\pi}{4}}
 v^2_0
 \left[
 v_0 {\rm Erf}\left(\frac{v_{\rm esc}}{v_0}\right)
   - \sqrt{2} v_{\rm esc} \exp\left(-\frac{v^2}{v^2_0}\right)
\right]~,\\
\eea
where, in this subsection, we take the circular speed $v_0$ (= $\sqrt{2/3} \ \times$ dispersion speed) = 220~km~s$^{-1}$~\cite{Smith:1988kw}, and the Milky Way escape speed
$v_{\rm esc}$ = 600~km~s$^{-1}$ unless specified otherwise.
This distribution is predicted by the Standard Halo Model~\cite{Drukier:1986tm}.
\\

{\bf \em Dispersion speed.}

Our method of extraction of the dispersion speed is as follows.
As argued in Sec.~\ref{sec:reconstruct} the fractional resolution in speed is too small for smearing to be important, nevertheless for completeness we obtain a smeared speed distribution, $f_{\rm recon} (v)$, by convolving Eq.~\eqref{eq:fMB} with the Gaussian function $(1/\sqrt{2 \pi \delta v})\exp(-v^2/2\delta v^2)$.
Next we draw $N_{\rm DM}$ events from $f_{\rm recon} (v)$, which constitutes the pseudo-data of our experiment.
For the purposes of generating pseudodata, we sample from a flux-weighted distribution $\propto v f_{\rm recon}$, {\em i.e.} from the expected detection rate of events with speed $v$.

We then determine how many events are required to faithfully extract the mean speed $v_{\rm mean}$, which can be easily computed from a set of sample events (and denoted by $v^{\rm recon}_{\rm mean}$) after flux-unweighting them.
The mean speed can then be compared to the mean speed predicted by Eq.~\eqref{eq:fMB}, which is given by
\bea
\nn v_{\rm mean} = \sqrt{\frac{4}{\pi}} \frac{v_0(e^{v^2_{\rm esc}/v^2_0}-1) - v^2_{\rm esc}/v_0}{e^{v^2_{\rm esc}/v^2_0}{\rm Erf}(v_{\rm esc}/v_0)-\sqrt{2}}~.\\
\label{eq:vmean}
\eea

For $v_0$ = 220~km~s$^{-1}$ and $v_{\rm esc}$ = 600~km~s$^{-1}$, we have $v_{\rm mean}$ = 249~km~s$^{-1}$.
To gauge the efficacy of our extraction of $v_{\rm mean}$ we repeat this procedure over 300 trials.
In the top left panel of Fig.~\ref{fig:vdisp} we show our trial distributions of the quantity
$(v^{\rm recon}_{\rm mean} - v_{\rm mean})/v_{\rm mean}$
for $N_{\rm DM} = 10^4, 10^3, 10^2$, after setting $v_{\rm esc}$ = 600~km~s$^{-1}$ in Eq.~\eqref{eq:vmean}.
Clearly $v_{\rm mean}$ is reconstructed better with better statistics.
In particular, we see that, with as few as $10^2$ events, $v_{\rm mean}$ can be reconstructed with a precision of $\sim 10\%$.  With
$10^4$ events, a reconstruction precision of $\sim 1\%$ is achievable.
Much larger exposures are required before the
speed resolution for individual dark matter particles has
a non-negligible effect on the precision with which $v_{\rm mean}$ is
reconstructed.

Strictly speaking, in the above procedure we must generate pseudodata by sampling events from a distribution in the Earth's rest frame, because
it is the flux in the Earth's rest frame which determines the velocity-distribution of events.
However, we have used the galactic frame distribution so that the relationship between $v_{\rm mean}$ and $v_0$ is simple as in Eq.~\eqref{eq:vmean}.
Moreover, the precision of reconstruction is not sensitive to the frame we pick, as borne out by the following hypothesis test.

One may wish to determine the number of events to be collected to reject some hypothesis for what $v_0$ is\footnote{There are some recent indications that the Milky Way's dark matter halo distribution may skew towards lower speeds than those found by fitting stellar velocity data~\cite{Herzog-Arbeitman:2017fte}.}.
To do so, we perform a Kolmogorov-Smirnnov~(\acro{ks}) test to differentiate between speed distributions for various~$v_0$, for which it will be useful to consider the expected flux-weighted distribution
of dark matter particle speeds in the Earth's rest frame~\cite{Gould:1987ir} (taken to be the Sun's rest frame):
\bea
\nn && v f_\oplus(v) \propto
v^2 \exp\left(\frac{-(v^2+v^2_\oplus)}{v^2_0}\right) \times \\
\nn & & \left[\exp\left(\frac{2vv_\oplus}{v^2_0}\right) - \exp\left(c_{\rm min}\frac{2vv_\oplus}{v^2_0}\right) \right] \Theta(v_{\rm esc} + v_\oplus - v)~,\\
\label{eq:vear}
\eea
where $v_\oplus = 235$~km~s$^{-1}$ is the mean Earth's speed in the galactic rest frame, and $c_{\rm min} = \max[-1,(v^2-(v^2_{\rm esc}-v_\oplus^2))/2vv_\oplus]$.
We plot the distribution $f_\oplus(v)$ for $v_0$~=180, 220, 260~km~s$^{-1}$ in the top right panel of Fig.~\ref{fig:vdisp}.

The results of our \acro{ks} test are shown in the bottom panels of Fig.~\ref{fig:vdisp}, where various $v_0$ are tested against a hypothesis of 220~km~s$^{-1}$.
We perform this test with 1000 trials and display the median result as well as the upper and lower quartiles.
We see that roughly 10 events suffice to determine $v_0$ to within $50$ km~s$^{-1}$, and $\Oc$(100) events will be required to determine it to within $10$ km~s$^{-1}$.
Thus our results from the \acro{ks} test are consistent with our results from reconstructing the mean speed: with ${\cal O}(10^2)$ events, the dark matter mean/dispersion speed can be determined to within a precision of $\sim 10\%$.

It is no surprise that the precision with which we can reconstruct the mean speed does not depend
dramatically on whether we used a speed distribution in the reference frame of the Earth, or in the galactic rest frame.  
Although
the galactic frame flux distribution (obtained from Eq.~\eqref{eq:fMB}) and the Earth frame flux distribution (Eq.~\eqref{eq:vear}) are
slightly different, their variances both scale as $v_0$.  
From the Central Limit Theorem, 
the precision with which the mean speed
can be reconstructed is determined only by the variance of the distribution and the number of events, irrespective of the detailed
shape of the distribution.  
More generally, this also indicates that our results should be fairly robust, even if the
speed distribution in galactic frame is not of the Maxwell-Boltzmann form.
\\

\begin{figure*}
\includegraphics[width=.95\textwidth]{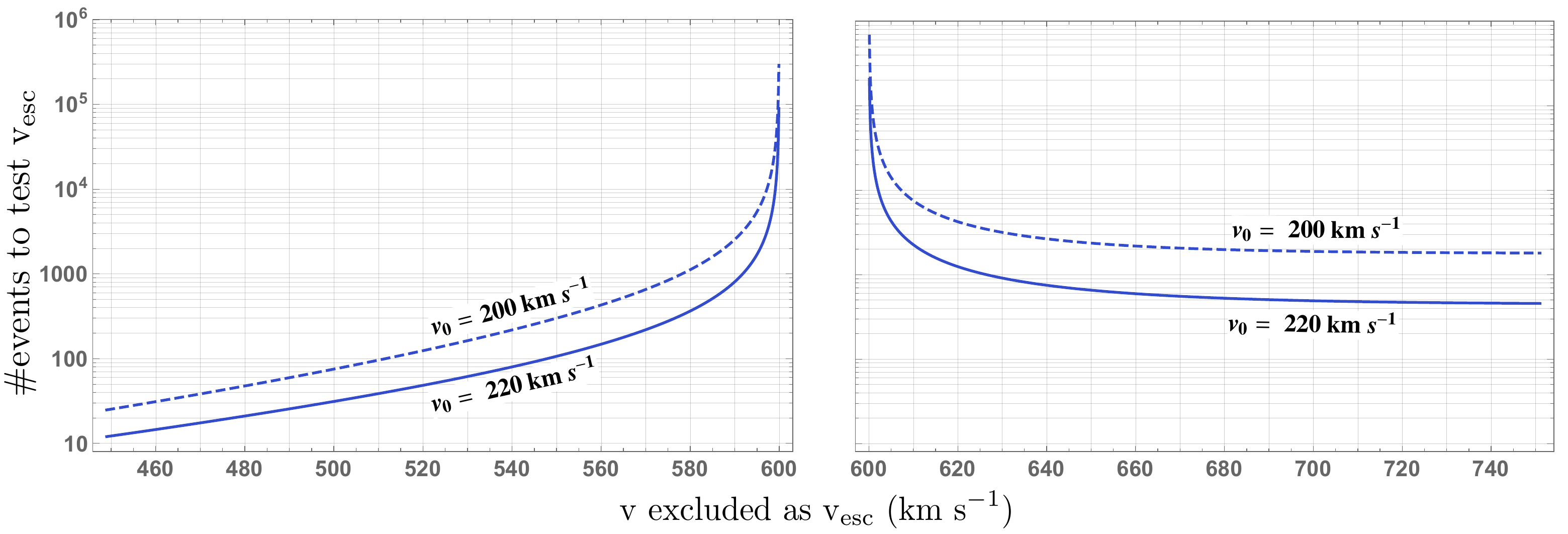}
\caption{The number of events required to exclude $v$ as the local escape speed of the galactic dark matter halo, assuming a truncated Maxwell-Boltzmann distribution with $v_0$ =200, 220~km~s$^{-1}$, and a null hypothesis with the true escape speed = 600~km~s$^{-1}$.
For $v <$ 600~km~s$^{-1}$, observing one event above $v$ would exclude it as the true escape speed.
For $v >$ 600~km~s$^{-1}$, observing zero events between 600~km~s$^{-1}$ and $v$, when 2.3 events are expected, would exclude it as the true escape speed at 90\%~\acro{c.l.} assuming Poisson statistics.
See Sec.~\ref{subsec:speedspec} for further details.
}
\label{fig:vesc}
\end{figure*}

{\bf \em Local escape speed.}

Next we estimate how sensitive large volume experiments will be to $v_{\rm esc}$.
At first glance it might appear easy to extract $v_{\rm esc}$ and $v_0$ from at least two moments of the distribution such as $v_{\rm mean}$ and, say, the root-mean-squared speed.
However, since $v_{\rm esc}$ by definition is a velocity feature that lies near the high speed tail of a Maxwell-Boltzmann distribution, these moments are very insensitive to $v_{\rm esc}$ and no faithful extraction is feasible. 
Furthermore, there is recent interest in whether a high-speed component of the local dark matter distribution may not follow Maxwell-Boltzmann statistics, having originated from the historic near passage of the Large Megallanic Cloud~\cite{Besla:2019xbx}.

Nevertheless it will be interesting to know how many events would need to be collected to test a truncated Maxwell-Boltzmann distribution, for the moment neglecting the possibility that the highest speeds of local dark matter may not follow this distribution.
For the sake of this test, we define the null hypothesis as specifying the escape speed to be $v^{\rm test}_{\rm esc}$~=~600~km~s$^{-1}$, which is near the median of local escape speeds determined by various recent surveys~
\cite{
RAVE,
SDSS,
Hattori:2018isv,
refId0}.
We can then exclude some $v < v^{\rm test}_{\rm esc}$ as the true $v_{\rm esc}$ if we observe a single event in the range [$v, \infty$], assuming the distribution in Eq.~\eqref{eq:fMB} with $v_{\rm esc}$ set to $v^{\rm test}_{\rm esc}$.
To do so, we assume that the actual number of events observed is exactly what we would expect from a Maxwell-Boltzmann distribution  truncated at $v_{\rm esc}^{\rm test}$ with $v_0$ as specified in Figure \ref{fig:vesc}.
Analogously, we can exclude some $v > v^{\rm test}_{\rm esc}$ as the true $v_{\rm esc}$ at 90\% \acro{c.l.} with Poisson statistics, if we expect 2.3 events in the range [$v^{\rm test}_{\rm esc}, v$],
a range in which 0 events will be observed.
Here we set $v_{\rm esc}$ in Eq.~\eqref{eq:fMB} to the $v$ we wish to exclude.
Of course, in practice the true escape speed -- the null hypothesis against which $v$ is excluded as $v_{\rm esc}$ -- will be unknown, therefore in the range $v > v^{\rm test}_{\rm esc}$, the quantity $v^{\rm test}_{\rm esc}$ must be interpreted as an accurately estimated Bayesian prior.
In Fig.~\ref{fig:vesc} we show the net number of events to be collected to satisfy these criteria as a function of $v$, for circular speeds $v_0$ = 200, 220~km~s$^{-1}$.
As expected, a higher $v_0$ implies that fewer events need to be collected to detect high-speed events.
We see that with $\sim 10^3$ events, the local escape speed can
be determined to within 20~km~s$^{-1}$, {\em i.e.} to within 3\%.
Once again due to the Central Limit Theorem, we expect a very similar precision had we carried out our test by drawing events from the Earth's rest frame.
Note that current uncertainties at the 90\%~\acro{c.l.} on the escape speed determined by astrophysical surveys range from 5\% to 10\%~\cite{
RAVE,
SDSS,
Hattori:2018isv,
refId0}.

For ease of comparison with existing bounds on $v_{\rm esc}$, we have assumed here that the velocity distribution of events is determined by the flux distribution
in galactic frame.  
A more accurate analysis could be made by drawing events from the flux distribution in Earth frame and then boosting back to
galactic frame, but as noted before, this will not change the analysis or the results qualitatively.
\\

{\bf \em Background rejection.}

Large volume cosmic particle detectors may identify a set of slow-moving candidate events in the coming years.
We show that the distribution of speeds in these events will provide a powerful background rejection method in the search for multiscattering dark matter, given the exquisite timing/speed information available at these experiments.

\begin{figure}
\includegraphics[width=.45\textwidth]{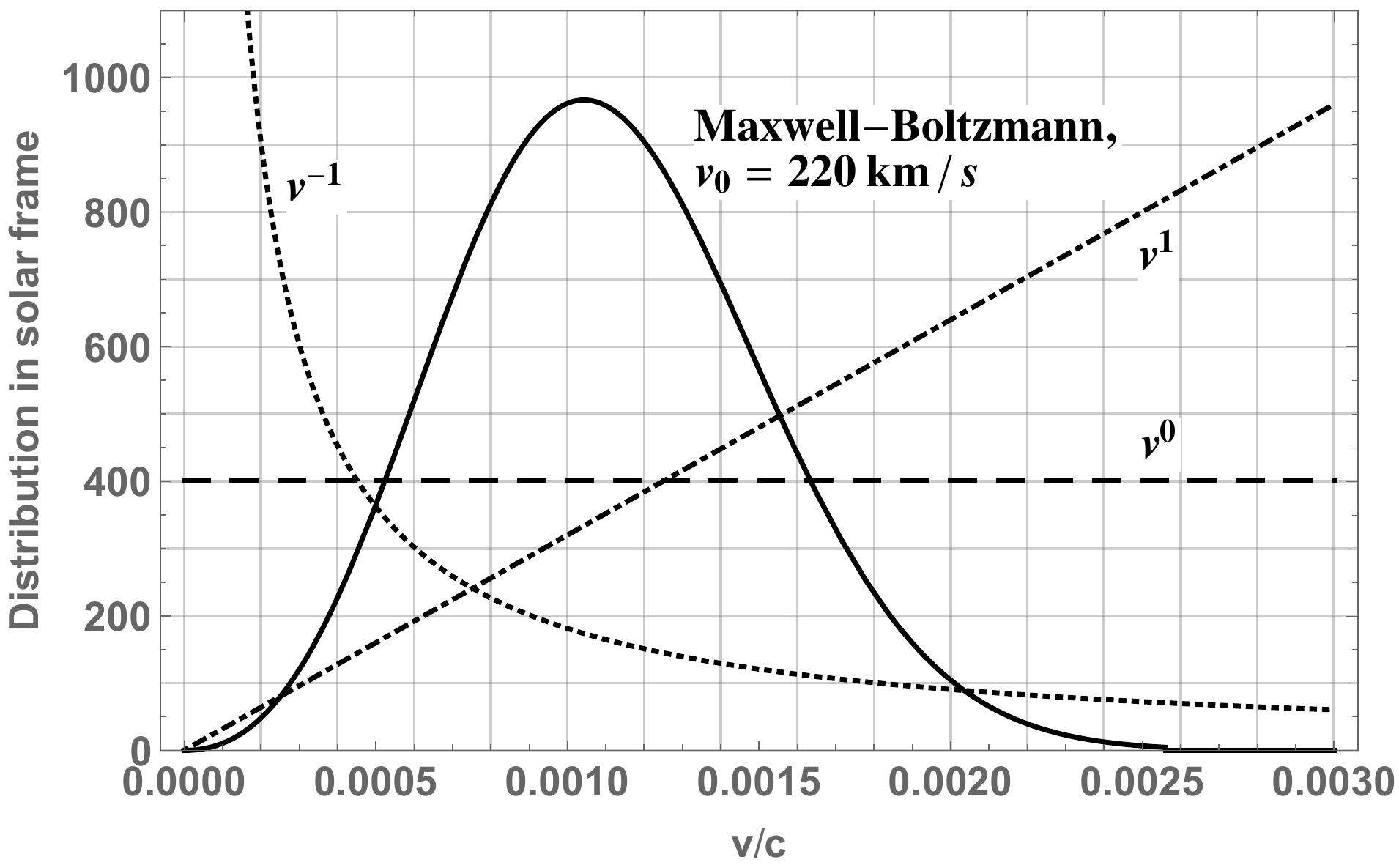} \\
\includegraphics[width=.44\textwidth]{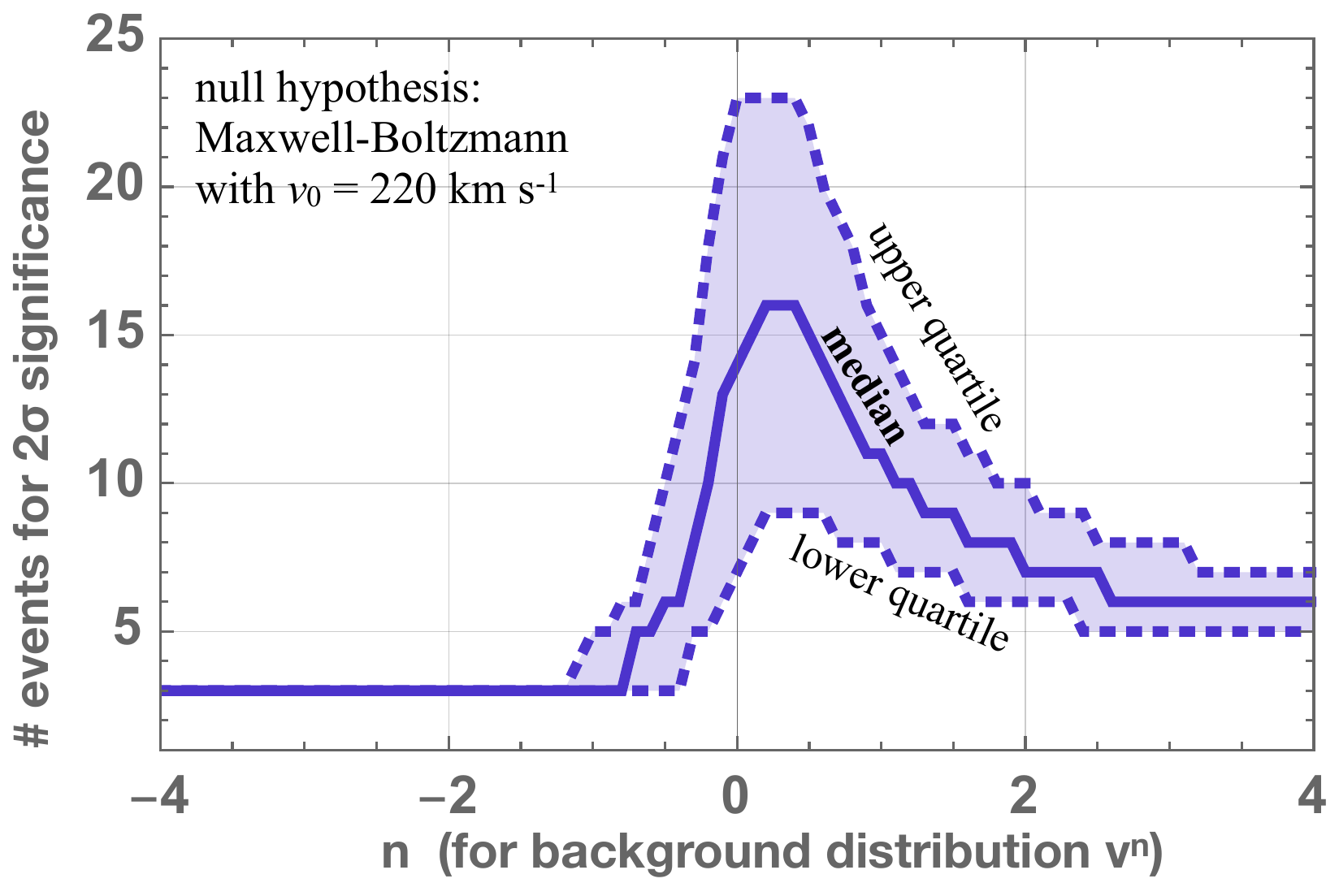}
\caption{{\bf Top}: A dark matter halo distribution in the Earth's rest frame $f_\oplus(v)$ taken from Eq.~\eqref{eq:vear} with circular speed $v_0 =220$ km~s$^{-1}$, and power-law background distributions normalized over the speed interval $v = 3- 750$ km~s$^{-1}$.
{\bf Bottom}: Results from 10$^3$ Kolmogorov-Smirnov trials determining the number of events required to reject with 2$\sigma$ significance background distributions $\propto v^n$ .
The dashed lines denote the upper and lower quartile of trials, the solid line denotes the median trial.
See Sec.~\ref{subsec:speedspec} for further details.
}
\label{fig:bkgd}
\end{figure}

Without having conducted a multiscatter dark matter event search, it is difficult to say what backgrounds may arise in the Earth's rest frame.
If these background events are indistinguishable from dark matter events, they may scale as some power of the background particles' speeds.
Specifically, we consider the power-law background distribution
\begin{align}
f_{\rm bkgd} (v) \propto v^n~,
\end{align}
 normalized over the range $v = 3- 750$ km~s$^{-1}$; it is necessary to truncate this range at the lower end for $n < 0$ since $f_{\rm bkgd} (v)$ is unbounded as $v\ra0$.
Figure~\ref{fig:bkgd} displays some normalized power-law background distributions alongside a dark matter Maxwell-Boltzmann distribution $f_\oplus(v)$ with $v_0$ = 220~km~s$^{-1}$ in Eq.~\eqref{eq:vear}.
To distinguish the backgrounds from dark matter, we perform a \acro{ks} test with 1000 trials: we randomly generate pseudodata from the above speed distributions, and determine how many events would be required to reject the Boltzmann distribution hypothesis at 2$\sigma$ significance.
Our results are shown in Figure \ref{fig:bkgd} with the median, upper quartile and lower quartile of trials.
We see that fewer than 25 events are required to discriminate signal events at 2$\sigma$ significance against a power-law background.

\begin{figure*}
\includegraphics[width=.45\textwidth]{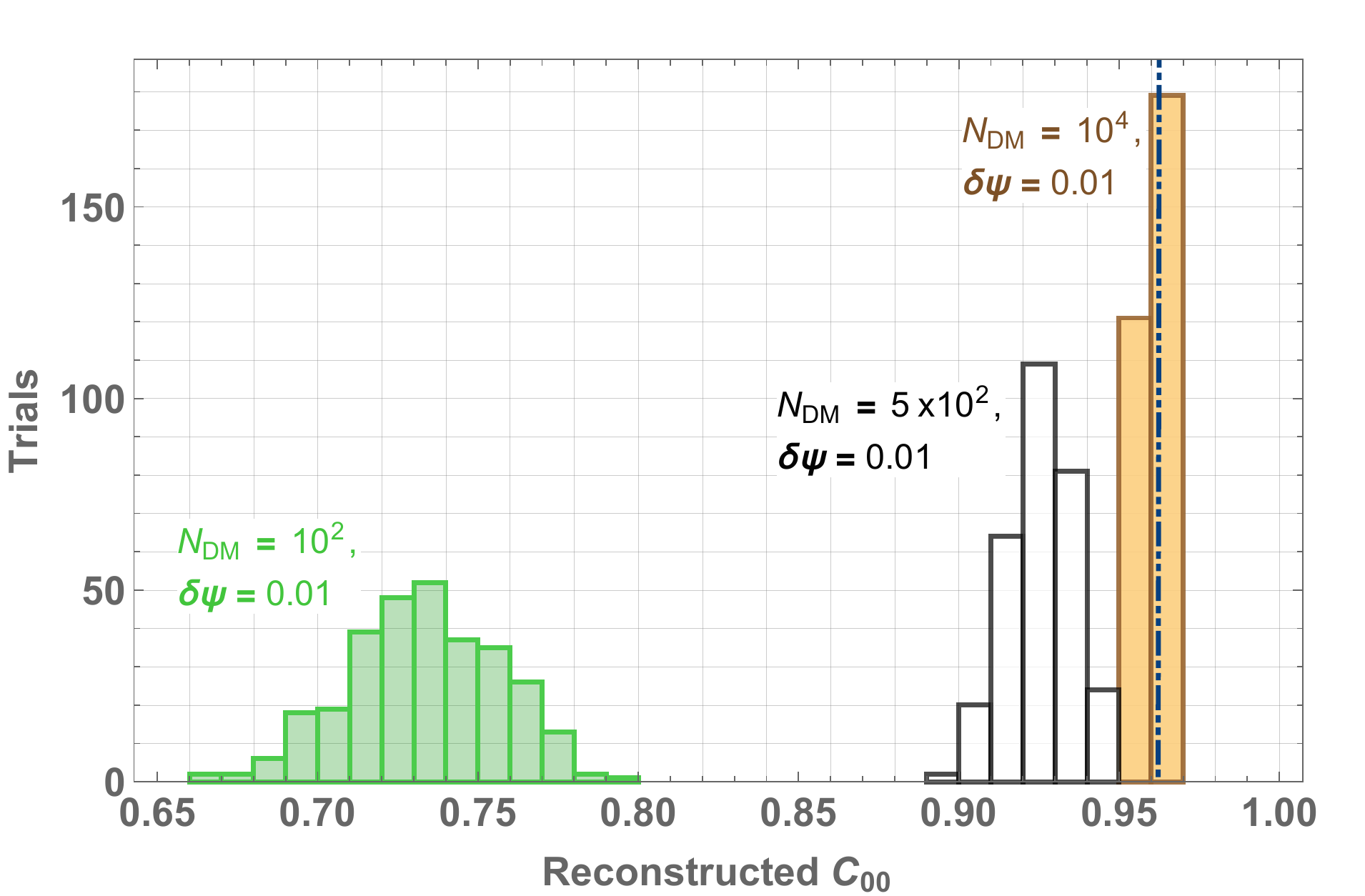}
\includegraphics[width=.45\textwidth]{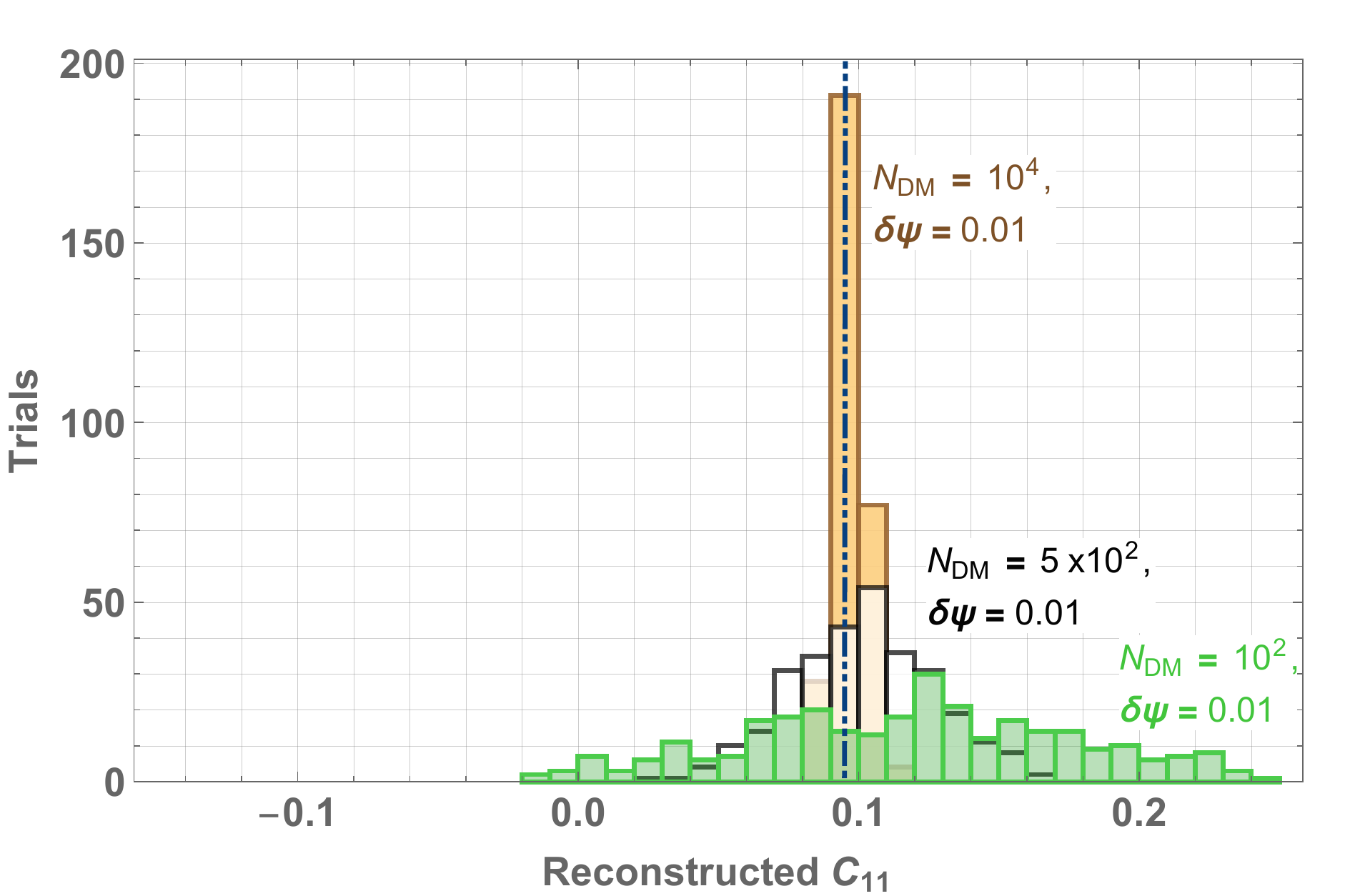} \\
\includegraphics[width=.45\textwidth]{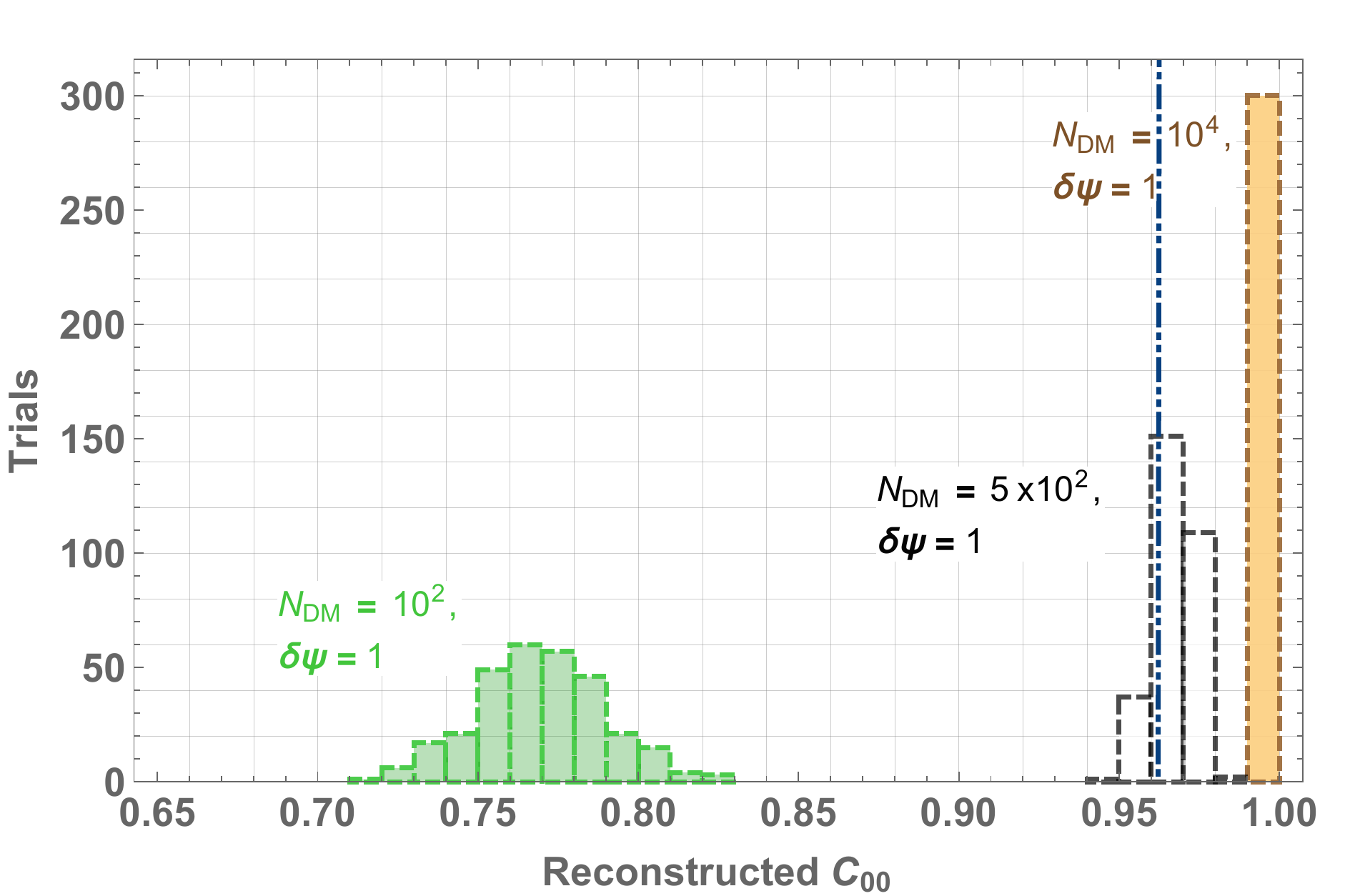}
\includegraphics[width=.45\textwidth]{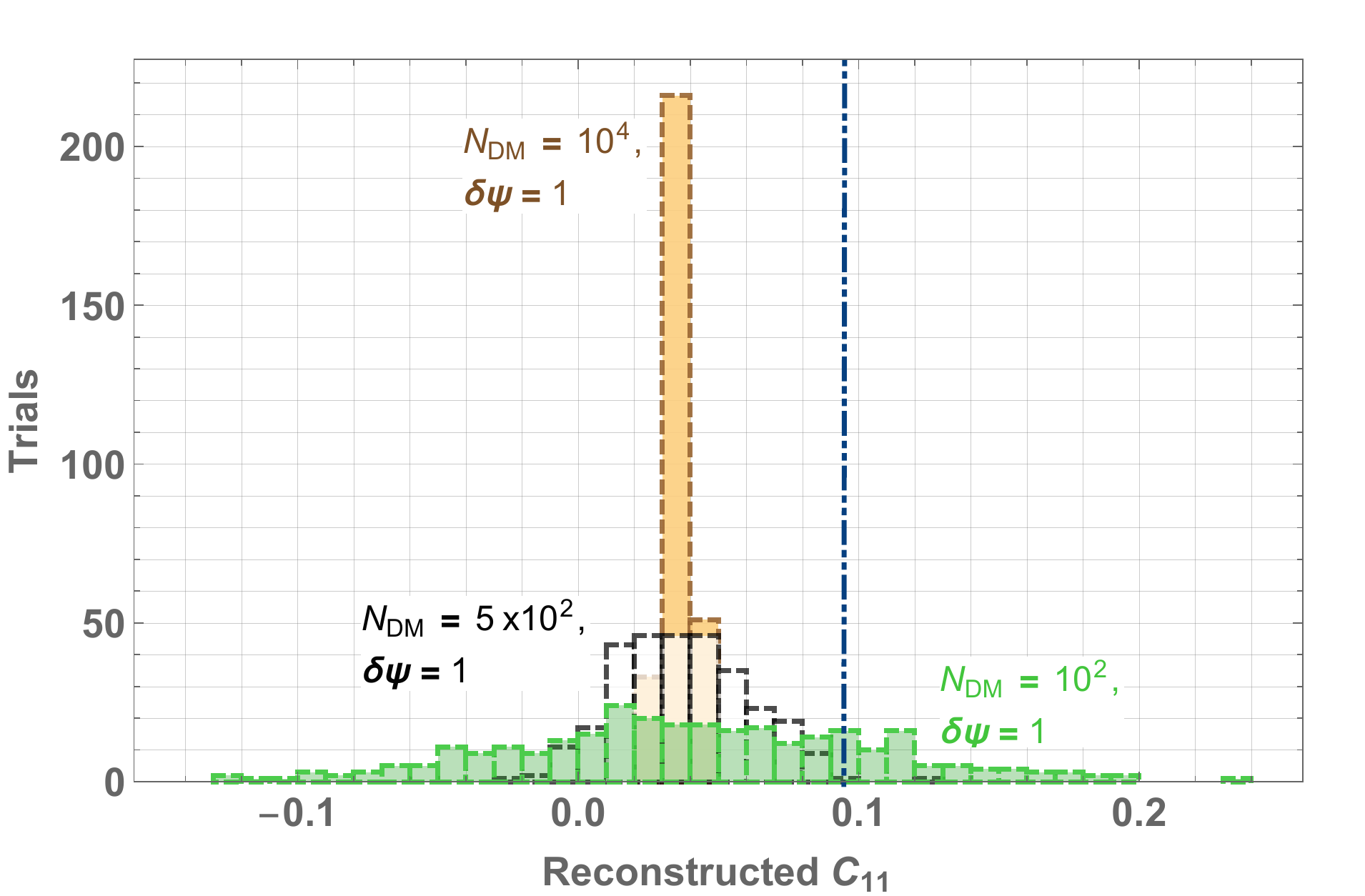}
\caption{
The effect of event statistics and detector resolution-induced smearing on the accuracy and precision of reconstructing the dark matter angular distribution.
Shown here are
distributions over 300 trials of the reconstructed coefficients of the monopole term  ($c_{00}$) and a dipole term ($c_{11}$) in the dark matter angular distribution assumed in Eq.~\eqref{eq:distang}, with 100, 500, and 10$^4$ events, for an angular resolution of 0.01 (top panels) and 1 (bottom panels).
The vertical dot-dashed lines indicate the actual values of these coefficients.
Greater statistics reconstruct the coefficients more accurately and precisely, and poorer resolutions wash out anisotropies.
See Sec.~\ref{subsec:angdist} for further details.
}
\label{fig:cxx}
\end{figure*}

\subsection{Angular distribution}
\label{subsec:angdist}

Dark matter velocities are usually thought to be isotropic in the galactic frame.
To see the effect of statistics and detector resolution on the reconstruction of this property, we introduce a ``test" anisotropy, and hence assume the (normalized) angular distribution of dark matter in the galactic frame is
\bea
\nn g(\theta, \phi)
&=& c_{00} Y_{00} + c_{\ell m} \sum_{\ell = 1,2} Y_{\ell m}   \\
\nn &=& \frac{1}{\sqrt{1+7\varepsilon^2}} \left( \sqrt{1-\varepsilon^2} Y_{00} + \varepsilon \sum_{\ell = 1,2} Y_{\ell m} \right)~,\\
\label{eq:distang}
\eea
where the $Y_{\ell m}$ are spherical harmonics in the real basis.
In the following we take $\varepsilon = 0.1$, so that in the above distribution the coefficient
\bea
 c_{\ell m} = \begin{cases} \sqrt{1-\varepsilon^2}/\sqrt{1+7\varepsilon^2} = 0.962; \ \ell = 0, m = 0~, \\
\ \ \ \ \ \ \ \ \ \varepsilon/\sqrt{1+7\varepsilon^2} = 0.097; \ \ell \neq 0~.
\end{cases}
\label{eq:clm}
\eea

Assuming that the uncertainties in polar and azimuthal directions are equal, {\em i.e.} $\delta \theta = \delta \phi = \delta \psi/\sqrt{2}$, we obtain the smeared distribution $g_{\rm recon}(\theta, \phi)$ by convolving the above function consecutively with the Gaussian functions $(1/\sqrt{\pi \delta\psi})\exp(-\psi_i^2/\delta\psi^2)$, with $\psi_i = \theta, \phi$.
We then draw $N_{\rm DM}$ events from $g_{\rm recon}(\theta, \phi)$ to obtain pseudo-data, and numerically reconstruct the coefficients $c_{\ell m}$ using the completeness relation.
We repeat this procedure over 300 trials.
In Fig.~\ref{fig:cxx} we show how the reconstructed values of $c_{00}$ and $c_{11}$ are distributed across these trials, for a realistic angular resolution of 0.01 (top panels) and a pessimistic one of 1 (bottom panels), for $N_{\rm DM} = 100,~500,~10^4$.
We also show with vertical dot-dashed lines the actual $c_{\ell m}$ values from  Eq.~\eqref{eq:clm}.
As expected, larger statistics reconstruct the coefficients more precisely; 10$^2$ events are sufficient for a reconstruct precision of $\sim$ 10\%, and 10$^4$ events for $\sim 1$\%.
But also interestingly, larger statistics reconstruct $c_{00}$ more {\em accurately}, identifying the isotropic nature of the distribution better;
we see that the reconstruction accuracy for $N_{\rm DM} = 100,~500,~10^4$ is $\sim 20\%,5\%,1\%$ respectively.
Finally, the larger the smearing, the nearer the co-efficient $c_{00}$ ($c_{11}$) is reconstructed to unity (zero), bearing out the intuition that smearing washes out anisotropies.
Although not plotted here, we find that the other $c_{\ell m}$ coefficients in Eq.~\eqref{eq:clm} exhibit reconstruction precisions very similar to that of $c_{11}$.

Note again that, for simplicity, we have performed our analysis in galactic frame.  In the frame of the solar system, the motion of the solar
system in the galactic plane induces an anisotropy in the angular distribution of events.  Our results thus indicate the extent to which one can distinguish
deviations from the anisotropies in Earth frame which would be expected from a distribution that is isotropic in galactic frame.

\section{DISCUSSION AND OUTLOOK}
\label{sec:conc}

While the discovery of dark matter certainly motivates searching for the unique  footprints left by multiscattering dark matter in timing data, the prospect of directly measuring dark matter halo properties at a terrestrial detector additionally inspires the program we have laid out.
In this study we have outlined how dark matter astrometry can be carried out at the currently operational liquid scintillator neutrino experiment \acro{sno}+, which could detect in 10 years up to 10$^5$ events of dark matter with a per-nucleon scattering cross section $\gsim$ 10$^{-28}$~cm$^2$, in a dark matter mass range of 10$^{16}-$10$^{21}$~GeV. 
We have shown that, thanks to \acro{sno}+'s superior timing resolution, the velocity vector of such multiscattering dark matter can be reconstructed for every event with an uncertainty so small as to render the effect of smearing from detector resolution negligible.
This reconstruction allows us to assemble an empirical velocity distribution, from which we could extract kinematic properties of the halo such as the dispersion speed, local escape speed, and velocity anisotropies, as well as reject environmental ({\em e.g.} instrumental, radiogenic or cosmogenic) backgrounds that may arise as power-law distributions by collecting just $\Oc$(10) events, using a statistical test such as the Kolmogorov-Smirnov test.

We have only attempted a basic estimate of the resolution with which a detector like \acro{sno}+ could pinpoint the speed and direction of a dark matter particle, on an event-by-event basis.  A more realistic estimate
would involve a detailed numerical attempt to model the detector response.  Although even our basic estimates indicate
that the uncertainty in particle direction is small, and the uncertainty in particle speed is negligible, it would
be interesting to verify these estimates with a more precise numerical study.

In discussing the prospects for liquid scintillator neutrino experiments to measure the local escape speed, we have neglected the
possibility that a population of dark matter particles is not bound to the Milky Way halo.  But if such a population were present and non-negligible,
then the techniques we have described provide a means for directly studying this population.  A detailed study of these prospects is beyond the scope
of this work, but would be an interesting topic for future investigation.

Our discussion of the reconstruction of dark matter tracks focused on \acro{sno}+, but it may be extended to other detector configurations as well, which we now remark on briefly.

{\bf \em Large volume neutrino detectors.}
Our reconstruction techniques can be easily applied to future liquid scintillator experiments, {\em e.g.} \acro{juno}~\cite{JUNO}, \acro{hanohano}~\cite{HANOHANO}, Deep-\acro{TITAND}~\cite{Deep-TITAND}, and \acro{mica}~\cite{MICA} once the detector configuration and timing resolutions become known.
Neutrino experiments that employ alternative detection technologies,
such as water \v{C}erenkov detectors ({\em e.g.} Super-K and Hyper-K),
liquid argon time projection chambers ({\em e.g.} \acro{dune}),
and optical modules ({\em e.g.} IceCube and \acro{antares}),
are unsuitable for this program because their energy thresholds are too high for detecting non-relativistic dark matter scattering.

{\bf \em Dark matter experiments.}
At noble liquid based detectors such as \acro{deap}-\osn{3600}~\cite{DEAP} where the dark mater signal is a single scintillation pulse, with reliance on pulse shape to discriminate from backgrounds, the velocity vector of multiscattering dark matter may be reconstructed in a manner very analogous to our description in Sec.~\ref{sec:reconstruct}.
At noble liquid time projection chambers such as \acro{xenon}-\osn{1}\acro{t}~\cite{X1T}, \acro{lux}~\cite{LUX}, PandaX-II~\cite{PANDAX}, and \acro{darkside}-\osn{50}~\cite{DARKSIDE}, where the dark matter signal comprises of two (scintillation + ionization) pulses, reconstruction of the interaction vertex, and thus of the velocity vector, is more challenging but feasible.
At \acro{darkside}-\osn{50} the liquid scintillator neutron veto system surrounding the inner detector may in principle be additionally deployed to detect multiscattering dark matter.
At bubble chamber detectors such as \acro{pico}-\osn{40}\acro{l}~\cite{PICO}, the dark matter direction may be easily obtained from a collinear trail of bubbles, but due to poor timing resolution the speed may be difficult to reconstruct.
We leave these challenges to the experimental collaborations to resolve, should they detect positive signals of multiscattering dark matter.

{\bf \em MATHUSLA.}
The recent proposal~\cite{MATHUSLA} to build a large volume hodoscope in the vicinity of a Large Hadron Collider general-purpose detector such as \acro{cms}, would also provide new sensitivity to multiscattering dark matter.
With an effective area of 100~m~$\times$~100~m, it would admit dark matter fluxes about 100 times higher than $\acro{sno}$+ and therefore probe dark matter masses proportionally greater.
M\acro{ATHUSLA} would contain multiple 1~cm-thick particle-tracking layers filled with extruded scintillator near the ceiling, and an additional tracking layer on the floor; the detection threshold of these scintillators is roughly 1 MeV.
Multiscattering dark matter would traverse each of these layers for $\Oc$(10)~ns, leave a collinear trail across layers, and transit the entire detector over $\Oc$(10) $\mu$s.
The most important background is cosmic ray showers at the rate of 1 shower per $\mu$s; vetoing this background using timing and tracking information requires further study.
Should isolation of the dark matter signature be achieved, it automatically reconstructs the dark matter velocity vector, and dark matter astrometry may be easily performed.
\\

In summary, we look forward not only to the discovery of dark matter at terrestrial detectors but also deploying them as dark matter speed guns and anemometers.

\section*{Acknowledgments}

We would like to thank
Mark Chen,
David Curtin,
Andrew B.~Pace,
Louis E.~Strigari,
Shawn Westerdale,
and
Alex Wright
for helpful conversations.
This work is supported by the Natural
Sciences and Engineering Research Council of Canada (\acro{nserc}).
T\acro{RIUMF} receives federal funding
via a contribution agreement with the National Research Council Canada.
The work of JK is supported in part by  Department of Energy grant DE-SC0010504.
For their hospitality, we are grateful to the organizers of the Mitchell Conference on Colliders, Dark Matter and Neutrino Physics 2019,
where part of this work was performed.
This work was also performed in part at the Aspen Center for Physics, which is supported by National Science Foundation grant PHY-1607611.


\appendix

\bibliography{darkastro}

\end{document}